\documentclass[11pt]{article}

\usepackage{amsfonts}
\usepackage{amsmath,amssymb}
\usepackage{bbm}
\usepackage{bm}
\usepackage{color}
\usepackage{slashed}
\usepackage{cite}
\usepackage[utf8]{inputenc}
\usepackage[toc,page]{appendix}
\usepackage{graphicx}
\usepackage{placeins}
\usepackage{caption}
\usepackage{subfigure}
\usepackage{booktabs}
\usepackage{comment}
\usepackage{physics}
\usepackage{float}
\usepackage{bbold}
\usepackage{bm}
\usepackage{subfigure}
\usepackage{siunitx}
\usepackage{todonotes}
\usepackage{dcolumn}
\DeclareGraphicsExtensions{.pdf,.png,.jpg}

\usepackage[colorlinks]{hyperref}
\AtBeginDocument{%
  \hypersetup{
    citecolor=blue,
    linkcolor=blue,   
    urlcolor=blue}}

\definecolor{airforceblue}{rgb}{0.36, 0.54, 0.66}
\definecolor{steelblue}{rgb}{0.27, 0.51, 0.71}
\definecolor{amber}{rgb}{1.0, 0.49, 0.0}
\definecolor{darkgreen}{rgb}{0.0, 0.5, 0.0}
\definecolor{amber}{rgb}{1.0, 0.49, 0.0}

\DeclareMathAlphabet{\mathpzc}{OT1}{pzc}{m}{it}

\graphicspath{{draft/Figures/}}

\allowdisplaybreaks[1]
\def\simg{{\ \lower-1.2pt\vbox{\hbox{\rlap{$>$}\lower6pt\vbox{\hbox{$\sim$}}}}\ }}
\def\siml{{\ \lower-1.2pt\vbox{\hbox{\rlap{$<$}\lower6pt\vbox{\hbox{$\sim$}}}}\ }}

\makeatletter \@addtoreset{equation}{section} \makeatother
\renewcommand{\theequation}{\arabic{section}.\arabic{equation}}

\newcommand{\geff}{g_{\text{eff}}}
\newcommand{\heff}{h_{\text{eff}}}
\newcommand{\gstar}{g_\star}
\newcommand{\vrel}{v_{\text{rel}}}
\newcommand{\kone}[1]{K_1\left(#1\right)}
\newcommand{\ktwo}[1]{K_2\left(#1\right)}
\newcommand{\ttilde}{\tilde{t}}
\newcommand{\mttilde}{m_{\tilde{t}}}
\newcommand{\deltamttilde}{\Delta m_{\ttilde}}
\newcommand{\tautilde}{\tilde{\tau}}
\newcommand{\mtautilde}{m_{\tilde{\tau}}}
\newcommand{\deltamtautilde}{\Delta m_{\tautilde}}
\newcommand{\mchi}{m_\chi}
\newcommand{\alphaem}{\alpha_{\text{em}}}
\newcommand{\alphasS}{\alpha^S_s}
\newcommand{\alphagSs}{\alpha_g^{[\mathbf{1}]}}
\newcommand{\alphagSo}{\alpha_g^{[\mathbf{8}]}}

\newcommand{\alphasBs}{\alpha^B_{s,[\mathbf{1}]}}
\newcommand{\alphagBs}{\alpha^B_{g,[\mathbf{1}]}}
\newcommand{\alphagB}{\alpha^B_{g}}
\newcommand{\alphasBSFs}{\alpha^{\text{BSF}}_{s,[\mathbf{1}]}}
\newcommand{\lchi}{\lambda_\chi}
\newcommand{\lH}{\lambda_H}
\newcommand{\Yttbar}{Y_{\ttilde\ttilde^*}}
\newcommand{\YB}{Y_{\text{B}}}
\newcommand{\phik}{\phi_{\vec{k}}}
\newcommand{\psinlm}{\psi_{nlm}}
\newcommand{\Ek}{\mathcal{E}_{\vec{k}}}
\newcommand{\En}{\mathcal{E}_{n}}

\renewcommand{\vec}[1]{\boldsymbol{#1}}

\begin{document}
\flushbottom

\begin{titlepage}

\begin{centering}

\vfill

{\Large{\bf Impact of bound states on non-thermal dark matter production}}

\vspace{0.8cm}

J. Bollig\footnote{julian.bollig@physik.uni-freiburg.de}
and S.~Vogl\footnote{ stefan.vogl@physik.uni-freiburg.de}

\vspace{0.8cm}

{Physikalisches Institut, Albert-Ludwigs-Universität Freiburg \\
Hermann-Herder-Str. 3, 79104 Freiburg,
Germany}
\vspace*{0.8cm}

\end{centering}
\vspace*{0.3cm}
 
\noindent

\textbf{Abstract}:  
We explore the impact of non-perturbative effects, namely Sommerfeld enhancement and bound state formation, on the cosmological production of non-thermal dark matter. For this purpose, we focus on a class of simplified models with t-channel mediators. These naturally combine the requirements for large corrections in the early Universe, i.e. beyond the Standard Model states with long range interactions, with a sizable new physics production cross section at the LHC. We find that the dark matter yield of the superWIMP mechanism is suppressed considerably due to the non-perturbative effects under consideration in models with color-charged mediators. In models with only electrically charged mediators the impact of non-perturbative effects is less pronounced and gets eclipsed by the impact of a possible Higgs portal interaction. In both cases we find significant shifts in the cosmologically preferred parameter space of non-thermal dark matter in these models. We also revisit the implications of LHC bounds on long-lived particles associated with non-thermal dark matter and find that testing this scenario at the LHC is a bigger challenge than previously anticipated.  
\vfill

\end{titlepage}
\setcounter{footnote}{0}

\section{\label{sec:intro}Introduction}

Dark matter (DM) is one of the most important puzzles of modern physics. Despite overwhelming evidence for its existence and a percent level determination of its cosmological abundance by the Planck collaboration,  $\Omega_{\mbox{\tiny DM}} h^2= 0.12 \pm 0.0012$ \cite{Planck:2018vyg}, neither its origin nor its nature is known. 
A promising avenue towards a better understanding is to leverage theoretical insights from the production of DM in the early Universe to make predictions for the expected signature in experiments. This approach is well established and has reached a considerable level of sophistication for DM candidates that are produced by thermal freeze-out, see e.g. \cite{Bertone:2004pz,Arcadi:2017kky} and references therein. An intriguing alternative, which has not reached a comparable amount of attention to date, is non-thermal production \cite{covi1999,Feng2003,Hall2010}. In this scenario, DM is characterized by very weak interactions with the Standard Model (SM) plasma such that it never reaches thermal equilibrium. Interestingly, this tends to make other states in the dark sector potentially long-lived which leads to very distinct signatures at collider experiments provided that these states can be produced efficiently.

In this paper, we focus on a minimal class of models that simultaneously allow for non-thermal DM and predict an appreciable production of new physics at collider experiments. To be concrete, we consider the very weak coupling limit of t-channel mediator models which are popular as a tool for the interpretation of LHC searches for DM \cite{Chang:2013oia,An:2013xka,Garny:2014waa,Ibarra:2015nca,Papucci:2014iwa,DiFranzo:2013vra,Kahlhoefer:2017dnp,Arina:2020tuw,arcadi2021systematic}.
In these models, the mediators possess SM gauge quantum numbers and are, therefore, easily produced at the LHC provided that their mass does not exceed $\mathcal{O}(1) \mbox{ TeV}$. However, aside from the abundant production at colliders, the SM gauge interactions also give rise to large non-perturbative corrections to the annihilation rates of the mediators in the early Universe. While these effects have been investigated in DM freeze-out before \cite{vonHarling2014,Mitridate:2017izz,Harz2018,Biondini:2018ovz,Biondini:2018pwp,Biondini:2019int,Binder:2020efn,binder2021nonabelian,bottaro2021closing}, they have not been considered in the context of non-thermal production and it is not clear how large their impact on the cosmologically preferred parameter space is.

To close this gap, we perform a state-of-the-art computation for non-thermal DM production from freeze-in (FI) and the superWIMP (sW) mechanism paying close attention to non-perturbative effects, i.e. Sommerfeld enhancement (SE) and bound state formation (BSF). The obtained results are then combined with experimental and observational bounds from the LHC and cosmology to identify the viable region of parameter space.

This paper is structured as follows: In Sec.~\ref{sec:nonthermalDM} we introduce our model and the DM production mechanisms considered here. In the next section, we describe the effect of Sommerfeld enhancement and bound state formation on the Boltzmann equations that govern the dynamics in the early Universe and discuss the relevant rates. We use these results to identify the cosmologically preferred parameter space and confront it with limits from LHC searches for long-lived particles and BBN bounds in Sec.~\ref{sec:pheno}. Eventually, we summarize our conclusions in Sec.~\ref{sec:conclusions}. Additional details of our computation are presented in Appx.~\ref{app:freezein} to \ref{app:QEDcase}.


\section{\label{sec:nonthermalDM} Non-thermal dark matter}


\subsection{\label{subsec:simplifiedmodel} The simplified model}
We consider a minimal simplified model with two new particles,  a DM candidate and a mediator with the SM. To allow for non-thermal production, the DM particle has to be a total singlet under the SM gauge group. We also assume that it is fermionic since this naturally avoids the presence of a Higgs portal connecting DM to the SM directly. 
To guarantee stability of the DM candidate and forbid coupling to the neutrino portal operator $LH$, we further assume a $\mathbb{Z}_2$ symmetry under which all (non-)SM particles are considered even (odd). This prevents DM decay to just SM particles. 
In the following we work with a Majorana fermion but the results for a Dirac fermion will be similar. The mediator is a scalar $\varphi$  which links the DM to the SM by a Yukawa interaction $\mathcal{L}\subset \lambda_\chi \bar{f} \varphi \chi $ where $f$ is a SM fermion and $\lambda_\chi$ the coupling. Gauge invariance of this interaction forces the quantum numbers of the scalar mediator to match those of the involved SM fermion. In principle, interactions with all SM fermions are possible. For simplicity, we focus on mediators that are a singlet under $SU(2)_L$. This restricts the interactions to right-handed SM fermions only and leaves two main possibilities: interactions with quarks, which require a color-charged mediator and interactions with charged leptons that only require hypercharge for $\varphi$. Since the flavor of the fermion will not affect the qualitative features of our scenario we restrict ourselves to two exemplary cases and only consider interactions with the top quark and the $\tau$ lepton. In order to distinguish these cases in a transparent way we now introduce a separate notation for each of them. Borrowing from supersymmetry we denote our top-philic color-charged mediator $\ttilde$ while the hyper-charged lepto-philic scalar is represented by $\tilde{\tau}$. An overview of their characteristics is given in Table \ref{tab:fields}. For the color-charged mediator the Lagrangian reads
\begin{eqnarray}
	\mathcal{L}_{\text{DS}}&=& i\bar{\chi}\gamma^\mu\partial_\mu\chi-\frac{1}{2}\mchi\bar{\chi}\chi-\mttilde^2\ttilde^*\ttilde\\
	\mathcal{L}_{\text{int}}&=&\abs{D_\mu \ttilde}^2+\lchi \overline{t}_R\ttilde\chi+\lH \ttilde^*\,\ttilde\abs{\Phi}^2+h.c.
\end{eqnarray}
where $D_\mu$ denotes the covariant derivative and $\Phi$ is the SM Higgs doublet.
The interaction part of the Lagrangian includes the Yukawa interactions between SM fermions, the mediator, and the DM and the interaction of the mediator with the gauge fields. In addition, there is also a Higgs portal term that connects the mediator to a pair of Higgs fields. The strength of this interaction is controlled by $\lH$ which is a free parameter in our model. Thus, we have a total of four parameters in our model; two masses, $m_\chi$ and $m_{\tilde{t}}$, and two couplings, $\lambda_\chi$ and $\lH$. We only consider $m_{\tilde{t}}\geq m_\chi$ since an inverted mass hierachy leads to a stable color charged relic with vastly different phenomenological consequences. The Lagrangian for the $\tau$-philic mediator is given in Appx.~\ref{app:QEDcase}.
\begin{table}
    \centering
    \caption{Summary of all new fields introduced in the simplified models considered. Besides their displayed type and charges under the SM gauge group, all these particles are odd under an additional $\mathbb{Z}_2$ symmetry.\label{tab:fields}}
    \label{tab:particleoverview}
    \begin{tabular}{cccc}
    \toprule
        \vspace{0.05cm}
		new particles & type & $SU(3)_c\cross SU(2)_L\cross U(1)_Y$ \\
		\midrule
		$\ttilde$ & bosonic scalar & $(\mathbf{3},\mathbf{1},4/3)$  \\[0.1cm]
		$\tautilde$ & bosonic scalar & $(\mathbf{1},\mathbf{1},-2)$  \\[0.1cm]
		$\chi$ & Majorana fermion & $(\mathbf{1},\mathbf{1},0)$ \\[0.1cm]
		\bottomrule
    \end{tabular}
\end{table}


\subsection{\label{subsec:prodmechanisms} Production mechanisms}

For the non-thermal production of $\chi$, two conditions have to be met: First, the number density of $\chi$, $n_\chi$, after reheating is very small such that the initial density does not saturate the DM abundance. We will assume $n_\chi(T_{rh})=0$ in the following where $T_{rh}$ is the reheating temperature. Second, the production of DM from the thermal bath has to be slow on cosmological scales throughout the evolution of the universe, i.e.
\begin{align}
\label{eq:estimate_non_thermal}
\left.\frac{\langle \Gamma \rangle_{\text{tot}}}{H}\right|_{T_{\text{max}}} \ll 1 
\end{align}
where $\langle \Gamma \rangle_{\text{tot}}$
is the total thermally averaged production rate, $H$ the Hubble rate and $T_{\text{max}}$ the temperature at which the production of $\chi$ is maximal.  This condition ensures that $\chi$ never reaches thermal equilibrium and it imposes a tight upper limit on the strength of the coupling~$\lambda_\chi$. Provided the two-body decay of the mediator to DM and a SM particle is kinematically allowed, the implication of this condition for the model parameters can be estimated  using  $\langle \Gamma \rangle_{\text{tot}} \approx \Gamma_{\ttilde\to t\chi}$ and $T_{max}\approx 0.3 \, \mttilde$ where an explicit expression of $\Gamma_{\ttilde\to t\chi}$ is given in Eq.~\ref{eq:decayrate}. This yields $\lambda_\chi \ll 2 \times 10^{-9} \sqrt{m_{\tilde{t}}/\mbox{GeV}}$ for $\mttilde \gtrsim 500$ \mbox{GeV} if $\chi$ is not mass degenerate with $\tilde{t}$ (or $\tilde{\tau}$). In contrast, the mediator $\ttilde$ thermalizes quickly since its SM gauge couplings lead to production and annihilation rates that are much fast than the expansion of the Universe for $T \gtrsim \mttilde$ . 

In full generality, the evolution of DM and the mediator is described by a system of coupled Boltzmann equations~\cite{Garny:2017rxs}. However, due to the large hierarchy of the involved interaction rates, it is possible to reduce the complexity of the problem. First, processes involving a DM particle in the initial state are negligible since the abundance of $\chi$ is small compared to its equilibrium value during the relevant stage of the evolution of the Universe. Second, as long as the interaction rate of $\tilde{t}$ is large compared to the Hubble rate, we can drop the equation describing its evolution and directly insert its equilibrium distribution in the one for $\chi$. This leads to the Boltzmann equation for freeze-in from decays and scattering~\cite{Hall2010}
\begin{align}
    \dv{n_\chi}{t} + 3 H n_\chi= 2 \langle \Gamma_{\tilde{t}} \rangle n_{\tilde{t}}^{eq}+ 2 \langle \sigma \vrel \rangle_{ab \rightarrow c\chi } n_{a}^{eq} n_{b}^{eq}
\end{align}
where the factor of $2$ comes from summing over the mediator and its antiparticle. The first term on the right-hand side accounts for the decays of $\tilde{t}$ and is proportional to the thermally averaged decay rate $ \langle \Gamma_{\tilde{t}} \rangle$ while the second term comes from scattering processes that produce $\chi$. Here, a sum over all contributions to the thermally averaged scattering cross section $\langle \sigma \vrel \rangle_{ab \rightarrow c\chi }$ is implied but we only include channels where one of the $a$, $b$ or $c$ is $\ttilde$ and the other two are SM particles since other processes are suppressed by a higher power of the small coupling $\lambda_\chi$.  
A concrete list of the processes included in our computation is given in Appx.~\ref{app:freezein}. 
As is conventional, we remove the Hubble expansion term on the right-hand side by introducing the yield $Y_\chi=n_\chi/s$, where $s$ is the entropy density. The resulting equation for $\dd{Y_\chi}/\dd{t}$ is independent of $Y_\chi$ such that it can be integrated directly. After changing the independent variable from $t$ to $x=m_{\tilde{t}}/T$ one gets
\begin{equation}
    \label{eq:freezein}
	Y^{\text{FI}}_\chi(x)=2\left[Y^{\text{FI}}_{\chi,1\to 2}(x)+\sum_{2\to 2} Y^{\text{FI}}_{\chi,2\to 2}(x)\right]\,.
\end{equation}
We present explicit expressions for $Y^{\text{FI}}_{\chi,1\to 2}$ and $Y^{\text{FI}}_{\chi,2\to 2}$ in Appx.~\ref{app:freezein}. Thermal corrections, which are known to lead to an $\mathcal{O}(10\%)$ correction to the relic density \cite{Biondini:2020ric}, are neglected. This mechanism contributes to the DM yield most strongly for $T\approx\mttilde$ and becomes quickly inefficient for $T\ll\mttilde$ due to the exponential suppression of the equilibrium density $n_{\tilde{t}}^{eq}$ in this regime. In practice, we find that the freeze-in contributions from $T\lesssim \mttilde/10$ are smaller than the observational uncertainty on $\Omega_{\mbox{\tiny DM}} h^2$ and can be neglected\footnote{Note that the depletion of the mediator abundance between $x=10$ and freeze-out is entirely due to annihilations since their decay rate is too small to have an impact on their own or the dark matter abundance. After freeze-out, however, all mediators have to decay to dark matter eventually due to their finite lifetime. Since this contribution is independent of the width as long as it is non-zero it can compete with the freeze-in contribution for small enough values the coupling.}. For extremely feeble Yukawa couplings  the amount of DM produced through freeze-in is insufficient to account for the amount of DM observed today. However, there is another production mechanism which becomes important (and also dominant) within this regime as we will discuss in the following.

Freeze-in production breaks down when $\tilde{t}$ drops out of equilibrium. This is guaranteed to happen for $T\ll m_{\tilde{t}}$ since the gauge interactions of the mediator are no longer strong enough to maintain chemical equilibrium in this regime and $\tilde{t}$ undergoes a freeze-out process. Neglecting the decays of $\tilde{t}$ for the moment, the Boltzmann equation for $\tilde{t}$ in this regime reads \cite{Kolb1990}
\begin{equation}
	\label{eq:MFOwoSEwoBSF}
	\dv{\Yttbar}{x}=-\frac{1}{2}\xi_1(x)\expval{\sigma_\text{ann}\vrel}\left(\Yttbar^2-\Yttbar^{eq\,2}\right),
\end{equation}
where $\Yttbar=Y_{\ttilde}+Y_{\ttilde^*}$ is the total yield of both $\tilde{t}$ and its antiparticle and $\expval{\sigma_\text{ann}\vrel}$  denotes the thermally averaged annihilation cross section. The prefactor $\xi_1(x)=\sqrt{\pi/45}M_{\text{Pl}}\mttilde\gstar^{1/2}(x) x^{-2}$, where $M_{\text{Pl}}$ is the Planck mass and $\gstar$ is a temperature dependent degree of freedom parameter that is defined in Appx.~\ref{app:freezein}. In the absence of the feeble interactions with $\chi$ this would lead to a cosmologically unacceptable (color) charged relic that spoils Big Bang Nucleosynthesis (BBN). However, a non-zero value of $\lchi$ allows for the decay of $\ttilde$ which saves BBN. The $\chi$s produced in these decays contribute to the DM relic density. This production mode is known as the superWIMP  mechanism following \cite{Feng2003,Feng2003a}, see \cite{covi1999} for earlier work. We get one DM particle per frozen-out mediator and, therefore, we have $Y_\chi^{\text{sW}}(x)|_{x\rightarrow\infty}=\Yttbar(x_{p})$ where $x_{p}$ is a post-freeze-out value of $x$ defined by the requirement that the yield is stable against further annihilations after this point. Since the time scale for these decays to happen is irrelevant as long as observational constraints on the lifetime of $\tilde{t}$ are respected, the DM yield from this production mode is largely independent of the coupling $\lchi$. This is in contrast with the freeze-in mechanism which predicts $Y_\chi^{\text{FI}} \propto \lambda_\chi^2$.


\section{\label{sec:nonperturbeff}Non-perturbative effects on dark matter production}
In a hot plasma with $ \Lambda_{\text{QCD}}\ll T$ QCD is not confining such that gluons can mediate long-range interactions between the mediators (the QED potential predominantly seen by the lepto-philic mediator $\tautilde$ is long-range anyways). Therefore, non-perturbative effects can have a significant impact on the production rates of particles. Conveniently, the main effects can be separated into distortions of the wave functions of scattering states, known as the Sommerfeld effect, and the existence of bound states. The Schroedinger equation describing the interaction between $\ttilde$ and $\ttilde^*$ in a potential $V(\vec{r})$  splits into
\begin{subequations}
\begin{eqnarray}
\label{eq:SEscattering}&&\left[-\frac{1}{2\mu}\nabla^2+V(\vec{r})\right]\phik(\vec{r})=\Ek\phik(\vec{r})\\
\label{eq:SEbound}&&\left[-\frac{1}{2\mu}\nabla^2+V(\vec{r})\right]\psinlm(\vec{r})=\En\psinlm(\vec{r})
\end{eqnarray}
\end{subequations}
for continuum solutions $\phik(\vec{r})$, i.e. the scattering states, and discrete solutions $\psinlm(\vec{r})$ that correspond to bound states. 
This heuristic picture can also be derived from quantum field theoretical considerations in the non-relativistic limit, see e.g.~\cite{Petraki2015}. 
It is sufficient to consider the QCD potential for the color-charged mediator since the strength of the QCD potential is large compared to the ones generated by the photon, the Z boson, or the Higgs (see Appx.~\ref{app:bsfmassbound} for more details and Appx.~\ref{app:QEDcase} for a discussion of the modifications that become relevant if a color-charge is absent). At leading order in $\alpha_s=g_s^2/(4\pi)$, where $g_s$ the (running) strong coupling constant, the potential is  Coulomb-like \cite{Fischler:1977yf}. 
In the non-relativistic limit, its energy eigenstates  are $\Ek=k^2/(2\mu)$ and $\En=-\kappa^2/(2\mu n^2)$ where $\mu=\mttilde/2$ is the reduced mass and $\vec{k}=\mu\vec{\vrel}$ the average momentum transfer in scattering processes while $\kappa=\mu \alphagB$ refers to the Bohr momentum. 
Here, $\alphagB$ denotes the interaction strength of the QCD potential in the bound state color representation evaluated at $\kappa$, which is thus only implicitly defined. 
Considering two particles $\ttilde$ and $\ttilde^*$ in the (anti-) fundamental $\mathbf{3}$ ($\bar{\mathbf{3}}$) representation, the combined scattering state
$\mathbf{3}\otimes\bar{\mathbf{3}}=\mathbf{1}\oplus\mathbf{8}$ can either be a singlet or an octet. 
For the bound state $\mathcal{B}(\ttilde,\ttilde^*)$ only a singlet is possible because the octet state would generate a repulsive potential. 
Therefore, $\alphagBs=4/3\alphasBs$ with $\alphasBs=\alpha_s(\mu \alphagBs)$. With analytical solutions for $\phik(\vec{r})$ and $\psinlm(\vec{r})$ the Sommerfeld corrections as well as the BSF cross sections can be calculated analytically \cite{Harz2018}. 
We restricted ourselves to s-wave contributions of the scattering state and the ground state ($\{nlm\}\to\{100\}$) in the following. 

Including non-perturbative effects changes Eq.~\ref{eq:MFOwoSEwoBSF} in two important ways. First, the annihilation cross section receives a large correction from the Sommerfeld effect \cite{Hisano:2004ds}. Second, for a sufficiently strong and long-ranged interaction, $\tilde{t}- \tilde{t}^*$ pairs can form bound states. Treating these as a separate particle species leads to a separated Boltzmann equation for the bound state $\mathcal{B}(\ttilde,\ttilde^*)$ (indicated by the subscript $B$ in the following) that is coupled to the original equation for the scattering states. This system of equations can be written as
\begin{subequations}
\label{eq:BSFBoltzmannAll}
\begin{align}
    \dv{\Yttbar}{x}&=-\frac{1}{2}\xi_1(x)\expval{\sigma_{\text{ann}}\vrel}\left(\Yttbar^2-\Yttbar^{eq\,2}\right)-\frac{1}{2}\xi_1(x)	\expval{\sigma_{\text{BSF}}\vrel}\Yttbar^2+2\xi_2(x)\expval{\Gamma_{\text{ion}}}\YB\label{eq:BSFBoltzmannYttbar}\\
	\dv{\YB}{x}&=-\xi_2(x)\expval{\Gamma_{\text{dec}}}\left(\YB-\YB^{eq}\right)+\frac{1}{4}\xi_1(x)\expval{\sigma_{\text{BSF}}\vrel}\Yttbar^2-\xi_2(x)\expval{\Gamma_{\text{ion}}}\YB\label{eq:BSFBoltzmannYB}
\end{align}
\end{subequations}
where we introduced a new factor  $\xi_2(x)=\xi_1(x)/s(x)$. In the non-relativistic limit the equilibrium yields are 
\begin{subequations}
\begin{eqnarray}
	\Yttbar^{eq}(x)&=&\frac{45g_{\tilde{t}}}{2\sqrt{2}\pi^{7/2}\heff(x)}x^{3/2}e^{-x}\\
	\label{eq:YBeq}
	\YB^{eq}(x)&=&\frac{g_B}{2g_{\tilde{t}}}\frac{\heff(\frac{m_B}{\mttilde}x)}{\heff(x)}\Yttbar^{eq}\left(\frac{m_B}{\mttilde}x\right)\,,
\end{eqnarray}
\end{subequations}
where $m_B=2\mttilde+\mathcal{E}_1$ is the mass of the ground state and $\mathcal{E}_1=-4/9\,\mttilde(\alphasBs)^2$ its binding energy. For $x>1$ we can safely take $\heff(m_B x/\mttilde)/\heff(x)\approx 1$ in our calculations. We present the corrections to $\expval{\sigma_{\text{ann}}\vrel}$ as well as the decay rate of the bound state $\expval{\Gamma_{\text{dec}}}$, the bound state formation cross section $\expval{\sigma_{\text{BSF}}\vrel}$ and the corresponding ionization rate $\expval{\Gamma_{\text{ion}}}$ in the following two subsections.


\subsection{\label{subsec:annplusdecay} Sommerfeld corrected annihilation cross sections}
Including the Sommerfeld effect, the annihilation cross section in the non-relativistic limit is given by
\begin{equation}
    \expval{\sigma_{\text{ann}}\vrel}\approx \sum_i \sigma_{0}^i \expval{S_{\text{ann},0}^{[\hat{\mathbf{R}}_i]}}
\end{equation}
where $\sigma_{0}^i$ are the s-wave approximations of the contributing annihilation processes and 
\begin{equation}
    \label{eq:Sann0Ri}
    \expval{S_{\text{ann},0}^{[\hat{\mathbf{R}}_i]}}= \frac{x^{3/2}}{2\sqrt{\pi}}\int_0^\infty \vrel^2e^{-\frac{x\vrel^2}{4}}S_{\text{ann},0}^{[\hat{\mathbf{R}}_i]}
\end{equation}
denotes the thermally averaged Sommerfeld factor. For a Coulomb potential with interaction strength $\alpha$, the Sommerfeld factor is given by
\begin{equation}
    S_0(\zeta)=\abs{\phik(\vec{r})}_{\vec{r}=0}^2=\frac{2\pi\zeta}{1-e^{-2\pi\zeta}}.
\end{equation}
with $\zeta=\alpha/\vrel$. In QCD, $\alpha\equiv\alpha_g^{[\hat{\mathbf{R}}_i]}$ depends on the color representation of the final state $\hat{\mathbf{R}}_i$ of the processes considered. Annihilations into $Z+Z$, $H+H$, $t+\bar{t}$ and $W^++W^-$ are restricted to the singlet states with $\alphagSs=4/3\,\alphasS$ whereas $g+\gamma$, $g+Z$ can only exist in an octet state with $\alphagSo=-1/6\,\alphasS$ and  $\alphasS=\alpha_s(\abs{\vec{k}})=\alpha_s(\mu\vrel)$ the QCD coupling strength at the average momentum of the scattering state. The Sommerfeld factors for singlet and octet potentials are, thus,  \cite{DeSimone:2014qkh}
\begin{equation}
     S_{\text{ann},0}^{[\mathbf{1}]}=S_0\left(\frac{4}{3}\zeta_S\right),\quad S_{\text{ann},0}^{[\mathbf{8}]}=S_0\left(-\frac{1}{6}\zeta_S\right)
\end{equation}
with $\zeta_S=\alphasS /\vrel$. For processes like $\ttilde+\ttilde^*$ annihilating into $g+g$, where both color configurations are possible, we have to correct for the treatment of the initial states in the computation of  the matrix elements that enter $\sigma_0^i$. This can be done by weighting the Sommerfeld factors, which for these two channels yield
\begin{equation}
    S_{\text{ann},0}^{[\mathbf{1}]+[\mathbf{8}]}=\frac{2}{7}S_0\left(\frac{4}{3}\zeta_S\right)+\frac{5}{7}S_0\left(-\frac{1}{6}\zeta_S\right)\,,
\end{equation}
see e.g. \cite{DeSimone:2014qkh,ElHedri2017} for a derivation of the weight factors. 


\subsection{\label{subsec:BSFplusIon} Bound state: formation, ionization and decay}
The bound state formation cross section to lowest order can be sufficiently described by radiating off an extra gluon $(\ttilde + \ttilde^*)_{[\mathbf{8}]}\to \mathcal{B}(\ttilde,\ttilde^*)_{[\mathbf{1}]}+g_{[\mathbf{8}]}$. Its cross section is given by \cite{Harz2018}
\begin{equation}
	\label{eq:sigmaBSFQCD}
    \sigma_{\text{BSF}}\vrel\approx\frac{2^717^2}{3^5}\frac{\pi\alphasBSFs\alphasBs}{\mttilde^2}S_{\text{BSF}}(\zeta_S,\zeta_B)
\end{equation}
where $\alphasBSFs=\alpha_s(\omega)$ is the QCD coupling strength at the gluon energy $\omega=\Ek-\En=\mu/2(\vrel^2+(\alphagBs)^2)$, $\zeta_B=\alphagBs/\vrel$ and 
\begin{equation}
	S_{\text{BSF}}(\zeta_S,\zeta_B)=\frac{2\pi\zeta_S}{1-e^{-2\pi\zeta_S}}\frac{(1+\zeta_S^2)\zeta_B^4}{(1+\zeta_B^2)^3}e^{-4\zeta_S\arccot(\zeta_B)}.
\end{equation}
The thermal average then reads
\begin{equation}
    \label{eq:avgBSFcrossection}
    \expval{\sigma_{\text{BSF}}\vrel}\approx\frac{x^{3/2}}{2\sqrt{\pi}}e^{\frac{x}{4}(\alphagBs)^2}\int_0^\infty \dd{\vrel}\vrel^2f_g(\omega)\sigma_{\text{BSF}}\vrel
\end{equation}
where $f_g(\omega)$ is the gluon distribution function in the SM plasma. The ionization rate as the reverse process $ \mathcal{B}(\ttilde,\ttilde^*)_{[\mathbf{1}]}+g_{[\mathbf{8}]}\to(\ttilde + \ttilde^*)_{[\mathbf{8}]}$ is related through the principle of detailed balance and reads
\begin{equation}
    \label{eq:ionizationrate}
    \expval{\Gamma_{\text{ion}}}= \frac{n^{eq\,2}_{\tilde{t}}}{n_B^{eq}} \expval{\sigma_{\text{BSF}}\vrel}\,. 
\end{equation}
We have repeated the analysis for a model with a lepto-philic mediator $\tautilde\subset(\mathbf{1},\mathbf{1},-2)$. The underlying physics is similar but there are some differences in the interaction rates, which we discuss in Appx.~\ref{app:QEDcase}.

Due to the singlet nature of the bound state, its decay rate is given by 
\begin{equation}
    \Gamma_{\text{dec}}\approx \abs{\psi_{100}(\vec{r})}^2_{\vec{r}=0}\sum_i \sigma^i_{B,0}
\end{equation}
where $\abs{\psi_{100}(\vec{r})}^2_{\vec{r}=0}=\kappa^3/\pi=8\mttilde^3(\alphasBs)^3/(27\pi)$ is the expectation value of the ground state wave function at the origin, $\sigma_{B,0}^i$ the s-wave color singlet annihilation cross section and the sum over $i$ runs over all allowed final states for the decay. In the channel $g+g$ that allows annihilations into a color singlet and octet configuration,  $\sigma_{B,0}^i=2/7\,g_{\ttilde}^2/g_B\sigma_0^i$  since we need to correct for the different color sum and the different number of degrees of freedom, while in the pure singlet channels $\sigma_{B,0}^i=g_{\ttilde}^2/g_B\sigma_0^i$. Thermal averaging of $\Gamma_{\text{dec}}$ adds an additional multiplicative factor of $K_1(m_B/T)/K_2(m_B/T)$ but we neglect this correction since it is approximately one in the freeze-out regime $x\gtrsim 20$.


\section{\label{sec:pheno}Phenomenology}


\subsection{\label{subsec:DMabundance}Dark matter abundance} 
\begin{figure}
\centering
    \includegraphics[width=0.75\textwidth]{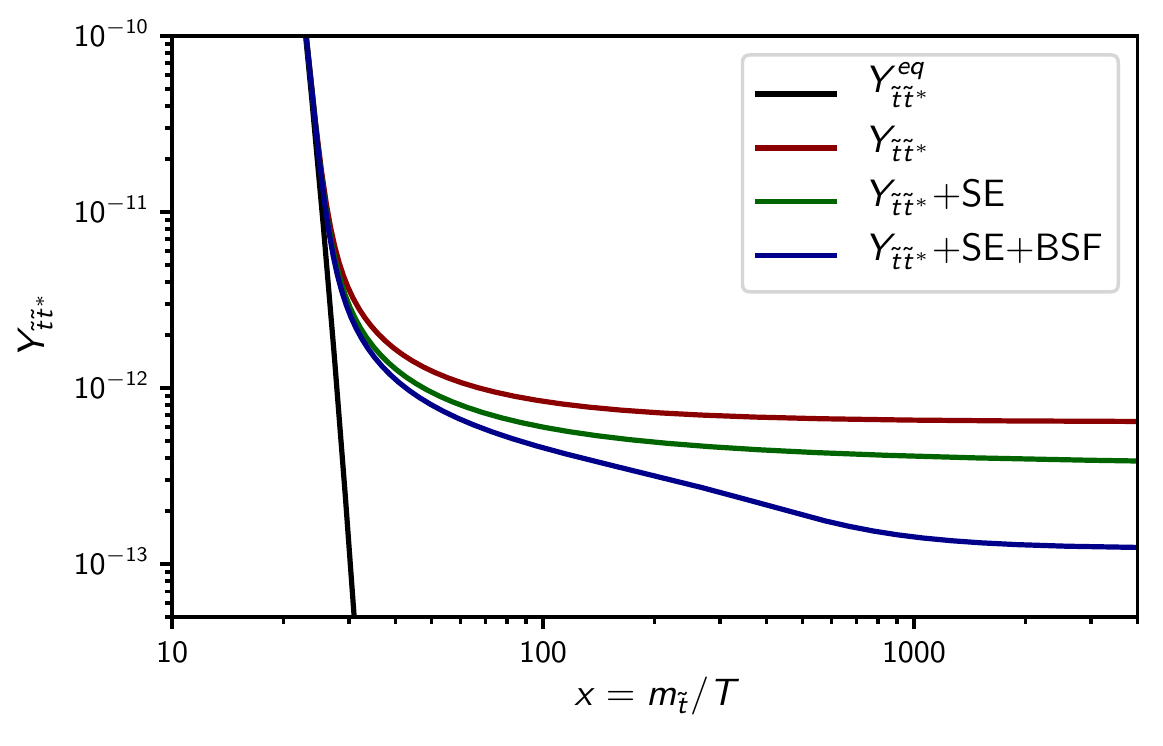}
    \caption{The yield for a color-charged mediator as a function of $x=\mttilde/T$ is shown for an exemplary mass of $\mttilde=\SI{5}{\TeV}$ and $\lH=0$. For comparison, the yield of a perturbative computation (red) is displayed alongside an SE corrected (green) and an SE and BSF corrected result (blue). The equilibrium yield is shown in black.}
    \label{fig:foexample}
\end{figure}

One can obtain the yield of the mediator $\Yttbar(x)$ in the superWIMP scenario by solving the Boltzmann equation(s). These types of equations do not have a known analytical solution and we employ numerical methods to solve them. However, the coupled system of equations in Eq.~\ref{eq:BSFBoltzmannAll} is still  challenging, especially in the region where the yields start to deviate from their equilibrium values at $x\approx 20$. Luckily, within this regime we can drastically simplify our calculations assuming Saha equilibrium, i.e. equilibrium between the scattering and the bound state. We have checked that this is well justified for $x\lesssim 30$. In Saha equilibrium $n_B/n_B^{eq}=n_{\tilde{t}}^2/n_{\tilde{t}}^{eq\,2}$ and thus $\Yttbar^2=\YB/\YB^{eq}\Yttbar^{eq\,2}$. For better numerical behavior, we also define $R(x)=n_B^{eq}/n_{\tilde{t}}^{eq\,2}$ such that $Y_B=R(x)s(x)\Yttbar^2$ and $\bar{x}=\YB/Y_{\text{tot}}$, i.e. the fraction of the bound state yield compared to the sum of the yields $Y_{\text{tot}}=\Yttbar+Y_B$ (similarly we have $\Yttbar^2/Y_{\text{tot}}^2=(1-\bar{x})^2$). Combining Eq.~\ref{eq:BSFBoltzmannYttbar} and \ref{eq:BSFBoltzmannYB} then leads to
\begin{equation}
    \label{eq:BSFBoltzmannSaha}
    \dv{Y_{\text{tot}}}{x}=-\frac{1}{2}\xi_1(x)\expval{\sigma_{\text{eff}}\vrel}\left((1-\bar{x})^2Y_{\text{tot}}^2-\Yttbar^{eq\,2}\right)
\end{equation}
with $\expval{\sigma_{\text{eff}}\vrel}=\expval{\sigma_{\text{ann}}\vrel}+2R(x)\expval{\Gamma_{\text{dec}}}$. As $\bar{x}\ll1$ we take $(1-\bar{x})^2\simeq 1$ and use  $Y(x_0)=\Yttbar^{eq}(x_0)$ at $x_0=1$ as our initial condition. We run Eq.~\ref{eq:BSFBoltzmannSaha} up to  $x= 30$ and  then use the obtained yields to initialize an ODE solver on the full system of equations given in Eq.~\ref{eq:BSFBoltzmannAll}. The results from this prescription agree with those obtained following the method proposed in \cite{Ellis:2015vaa,Harz2018} to $\lesssim 1\%$ throughout the parameter space considered here. This method can also be generalized to include transitions between bound states which are neglected here \cite{binder2021saha,garny2021bound}. For illustration, we show the solutions to the Boltzmann equation for an exemplary parameter point with $\mttilde=\SI{5}{\TeV}$ and $\lH=0$ in Fig.~\ref{fig:foexample}. The yield for the perturbative rates (red) is compared to a computation that includes only SE corrections to the cross section (green) and the full results which include both SE and BSF (blue). 
As can be seen, the formation of bound states has a significant impact on the mediator abundance $\Yttbar$, reducing it by roughly half an order of magnitude. This fits our qualitative expectations; the formation of bound states increases the effective cross section of the overall process and thus reduces the yield. The relatively efficient annihilations of a color-charged mediator at  $x=\mathcal{O}(100)$, which lead to the visible dip of the full result in this regime, have previously been observed in studies of coannihilation scenarios for freeze-out, see e.g. \cite{Harz2018,Biondini:2018pwp,Biondini:2018ovz}). The freeze-out process of the mediator particle is quite long and only comes to its end at $x\simeq 1000$. After this point the yields is stable and we can safely extract $\Yttbar(x_p)$ here. When the predicted relic density from the superWIMP mechanism $\Omega_{\text{sW}}= m_\chi s_0 \Yttbar(x_p) /\rho_{crit,0}$, where $\rho_{crit,0}$ is the critical energy density of our present day Universe, is larger than the observed value, the corresponding combination of $m_\chi,\, \mttilde$ and $\lH$ is excluded unless modifications of the subsequent cosmological history, e.g. late entropy production \cite{Co:2015pka,Calibbi:2021fld}, are considered. A too small $\Omega_{\text{sW}}$ can be made up by additional production from FI. Assuming that it makes up the rest we can define the required FI fraction $p=1-\Omega_{\text{sW}}/\Omega_{\text{\tiny DM}}$ which fixes $\lambda_\chi$ for a given $m_\chi,\, \mttilde$ and $\lH$ via
\begin{equation}
    \label{eq:lambdachi}
     \lchi=\sqrt{\frac{p\,\Omega_{\mbox{\tiny DM}}\rho_{crit,0}}{s_0Y_{\text{FI,norm}}^\infty\mchi}},
\end{equation}
where $Y_{\text{FI,norm}}^\infty=\lim_{x\to\infty}Y^{\text{FI}}_\chi(x)/\lambda_\chi^2$ is computed by pulling $\lambda_\chi^2$ out of Eqs.~\ref{eq:Yfi1to2} and just keeping the mass dependence. It can safely be evaluated at $x\gtrsim 10$.

For a lepto-philic mediator, BSF does not lead to an appreciable change of the yield compared to the SE corrected solution in the parameter space considered here. These observations are not surprising since the binding energy is much smaller here. Overall, the corrections from non-perturbative effects are $\lesssim 15\%$ such that we do not display these results graphically. 


\subsection{\label{subsec:constraints} Cosmological and laboratory constraints} 

There are various cosmological constraints. In the region of parameter space we are interested in, the most important limit comes from Big Bang Nucleosynthesis (BBN). The long-lived mediator particles can spoil it in two ways \cite{Jedamzik:2007qk,Kawasaki:2017bqm}. First, they can decay during the formation of the light elements and inject energy into the SM plasma. Second, if they are stable on time scales that are long compared to the duration of BBN they can form bound states with SM particles that can catalyze nuclear reactions that are not expected in the SM. Together, these effects upset the delicate balance of the involved nuclear reactions and change the abundances of the light elements. This can be avoided if the relics of the mediator $\tilde{t}$ (and $\tilde{\tau}$) either decay before the relevant stage of BBN or are initially produced with such a small yield that their presence does not affect the nuclear reaction appreciably. Given the expected yield from the bound state enhanced freeze-out we need to require that the lifetime $\tilde{\tau}=1/\Gamma_{\tilde{t}}\lesssim \SI{10}{\second}$ for the top-philic mediator \cite{Kawasaki:2017bqm}. If the two-body decay $\tilde{t}\rightarrow t_R + \chi$ is allowed the value of $\lambda_\chi$ preferred by the relic density constraint typically leads to sufficiently fast decay. However, when this channel closes the three-body decay $\tilde{t}\rightarrow W^+ +b+\chi$ becomes the dominant contribution to the width and the lifetime increases considerably for a fixed $\lambda_\chi$. We have computed $\Gamma_{\tilde{t}\rightarrow Wb\chi}$ numerically and include it in our analysis below the threshold of the two-body decay.  


A second class of cosmological constraints becomes relevant if the present day dark matter velocity is large enough to affect structure formation. Intuitively, this can be understood based on the free-streaming length, i.e.~the distance a dark matter travels after matter radiation equality. If this distance is larger than the size of the primordial density fluctuations the dark matter particles stream out of the overdensities and prevent the growth of structures on these scales.    
Traditionally, results for warm dark matter (WDM) are reported in terms of the mass $m_{\mbox{\tiny WDM}}$ of a dark matter particle with a thermal  velocity distribution. At the fundamental level this is better understood as a bound on the root-mean-square velocity of the DM today  which can be mapped to the  mass by the relation \cite{Bode:2000gq,Barkana:2001gr}
\begin{align}
     v_{rms}\approx 0.04 \left(\frac{\Omega h^2}{0.12}\right)^{1/3}  \left(\frac{m_{\mbox {\tiny WDM}}}{1\,\mbox{keV}}\right)^{-4/3}\frac{\mbox{km}}{\mbox{s}}\,.
\end{align}
Various astrophysical observations can be used to constrain $m_{\mbox{\tiny WDM}}$ 
and typically masses less than a few keV are excluded, 
see e.g. \cite{Irsic:2017ixq,Dekker:2021scf,Hsueh:2019ynk,Gilman:2019nap}. 
Taking the limit $m_{\mbox{\tiny WDM}} \geq 3.5$ keV from  \cite{Irsic:2017ixq}, which is based on an analysis of  Lyman-$\alpha$ forest data, leads to $v_{rms}\leq 7.5 \,\mbox{m/s}$.

Irrespective of the production mechanism, the velocity of a dark matter particle today is given by
\begin{align}
    \label{eq:v0WDMbound}
   v_0=\frac{p_{prod}}{m_\chi}\frac{a_{prod}}{a_0}=\frac{p_{prod}}{m_\chi}\left(\frac{g_0}{g_{d}}\right)^{1/3}\frac{T_0}{T_{prod}}
\end{align}
where $a$ is the scale factor and the subscript $prod$ (0) indicates qualities at the time of dark matter production (today).

For DM produced by the freeze-in mechanism  the mean momentum at production is $\mathcal{O}(1) T_{prod}$ provided $m_{\tilde{t}}\gg m_\chi+m_t$. Therefore, the dependence on $T_{prod}$ drops out of the equation and, taking the value $2.5$ for the $\mathcal{O}(1)$ factor as suggested by \cite{Heeck:2017xbu},
\begin{align}
    v_{0,FI}\approx\frac{2.5 T_0}{2 m_\chi}\left(\frac{g_0}{g_{d}}\right)^{1/3}\approx 30  \frac{1}{m_\chi[\mbox{keV}]} \frac{\mbox{m}}{\mbox{s}}\,.
\end{align}
 Comparing with the limit on $v_{rms}$ (a conversion of $v_{rms}=(\bar{v^2})^{1/2}$ to the mean velocity $\bar{v}$ only leads to a $10\%$ correction) yields $m_\chi \gtrsim 4$ keV. This is remarkably close to the limit of $15$ keV reported by \cite{Decant:2021mhj}  based on a full modeling of the transfer function for freeze-in. 
 
 A similar argument can be made for DM from the superWIMP mechanism. Here, the DM particle is produced by the decay of the frozen-out non-relativistic mediator and, therefore, the momentum at production is set by the kinematics of the decay.  Modeling the decay as instantaneous the temperature at production can be estimated by $\Gamma=H|_{T_{prod}}$, see e.g. \cite{Jedamzik:2005sx}.  In contrast to the freeze-in case the bound now depends on $m_\chi, m_{\tilde{t}}$ and $\lambda_\chi$ instead of $m_\chi$ alone. Assuming that all DM is produced through the superWIMP mechanism we can fix $m_\chi$ for a given $\mttilde$ as described in the previous section. Then, by employing Eq.~\ref{eq:v0WDMbound} with $p_{prod}\approx(\mttilde^2-m_\chi^2)/(2\mttilde)$ for a mediator decaying at rest (together with $g_d\sim 10$ and $v_0=7.5 \,\mbox{m/s}$ as an upper bound), one obtains a lower bound on $\lambda_\chi$ for each set of masses. 
 Using the full result for the relic density that we present in detail in the following section one finds
 $\lambda_\chi\gtrsim 10^{-15}$ for $\mttilde=5$ TeV 
 and values of $m_\chi$ that have an $\mathcal{O}(1)$ 
 contribution of the superWIMP mechanism to the relic density. This lower limits becomes tighter for higher masses but never exceeds a few times $10^{-14}$ for the masses considered here. Therefore, the warm dark matter bound does not provide a stringent constraint on the parameter space here.

Apart from cosmological constraints, there are also laboratory constraints that are relevant here. The most powerful tool for testing heavy new physics with experiments on Earth is the LHC. Provided that their mass is not too high, the mediator $\tilde{t}$ (or $\tilde{\tau}$) can be produced copiously in proton-proton collisions through Drell-Yan-like processes since they are charged under the SM gauge group\footnote{In principle, production via the Higgs is also possible. However, for reasonable values of $\lH$ we expect the cross section of this process to be small compared to the one due to gauge interactions \cite{Hessler2015}.}. In particular the color-charged mediator has a very high production cross section of $\sigma_{\ttilde\ttilde^*}\simeq\SI{1}{\femto\barn}$ for $\mttilde\sim\SI{1300}{\GeV}$ (an uncharged mediator has $\sigma_{\tautilde\tautilde^*}\simeq\SI{0.1}{\femto\barn}$ at $\mtautilde\sim\SI{500}{\GeV}$ for comparison). In addition, the condition for non-thermal production in Eq.~\ref{eq:estimate_non_thermal} tells us that we ought to expect $\lambda_\chi$ in the ballpark of $10^{-8}$ or lower. This does not only make the mediator long-lived compared to the Hubble time around the cosmological production of $\chi$ but also implies a long lifetime in the lab today. The expected decay length in the lab is given by
\begin{equation}
	\Delta x \simeq \frac{\hbar}{\Gamma_{\ttilde}} \beta\gamma c
\end{equation}
where  $\gamma$ is the Lorentz factor and $\beta$ the velocity of the decaying particle in natural units. 
We have checked that for the values of $\lambda_\chi$ preferred by freeze-in and with a realistic distribution of $\gamma$ factors $\Delta x \geq \SI{100}{\metre}$ throughout the considered parameter space. Therefore, the mediators are not just long-lived but can be treated as stable on the scales of collider experiments. Such massive, slow particles can be searched for at the LHC utilizing their unusual energy loss or with time-of-flight measurements \cite{ATLAS2019,CMS-PAS-EXO-16-036}. We take the most stringent limits on a stop-like R-hadron and the limits on a directly produced stau from \cite{ATLAS2019} based on an integrated luminosity of $\SI{36.1}{\femto\barn^{-1}}$ collected at $\sqrt{s}=13$ TeV.  The measurements exclude $\mttilde\lesssim\SI{1350}{\GeV}$ at $95\%$ CL. Assuming a similar efficiency and an integrated luminosity of $\SI{3000}{\femto\barn^{-1}}$, for the HL-LHC we expect the exclusion limits to rise to $\mttilde\approx\SI{1600}{\GeV}$ in the future. For a lepto-philic mediator, the production cross section is significantly lower and we find a lower limit of $\mtautilde\gtrsim \SI{430}{\GeV}$.  An estimate following the reasoning outlined above indicates that the full data set available at the end of the HL-LHC will exclude $\mtautilde\lesssim \SI{600}{\GeV}$. The phenomenology in these model benchmarks is very similar to the one expected in our case but it is not strictly the same. In particular, the production of supersymmetric particles can receive contributions from other superpartners that are absent in our model. Therefore, we have reevaluated the theoretical production cross section with CalcHEP \cite{Belyaev:2012qa}. We find good agreement between our results and the theory prediction reported in \cite{ATLAS2019} which points towards a limited influence of supersymmetry specific effects on the production and justifies using the limits on these models for our purposes.


\subsection{\label{subsec:parameterspace}The parameter space} 

Finally, we can put our results for the relic density and the observational limits together. Since $\Omega_{\mbox{\tiny DM}}$ is known very precisely, we can use it to fix one of the parameters of our model in terms of the others. We choose to keep the masses $\mttilde$ and $\mchi$ as well as $\lH$ and predict $\lambda_\chi$. The Higgs coupling is a free parameter that can be used to boost the annihilation cross section of the mediators and the decay rate of the bound state while leaving the non-perturbative effects and the freeze-in largely unaffected. We investigate its impact on the phenomenology by comparing the results for a minimal scenario with $\lH=0$ and one with $\lH=0.3$, which is a sizable value but not dangerously large.

\begin{figure}
    \centering
    \subfigure[Viable parameter space for a top-philic mediator model assuming $\lH=0$ (solid lines) and $\lH=0.3$ (dashed lines).]{\includegraphics[width=0.75\textwidth]{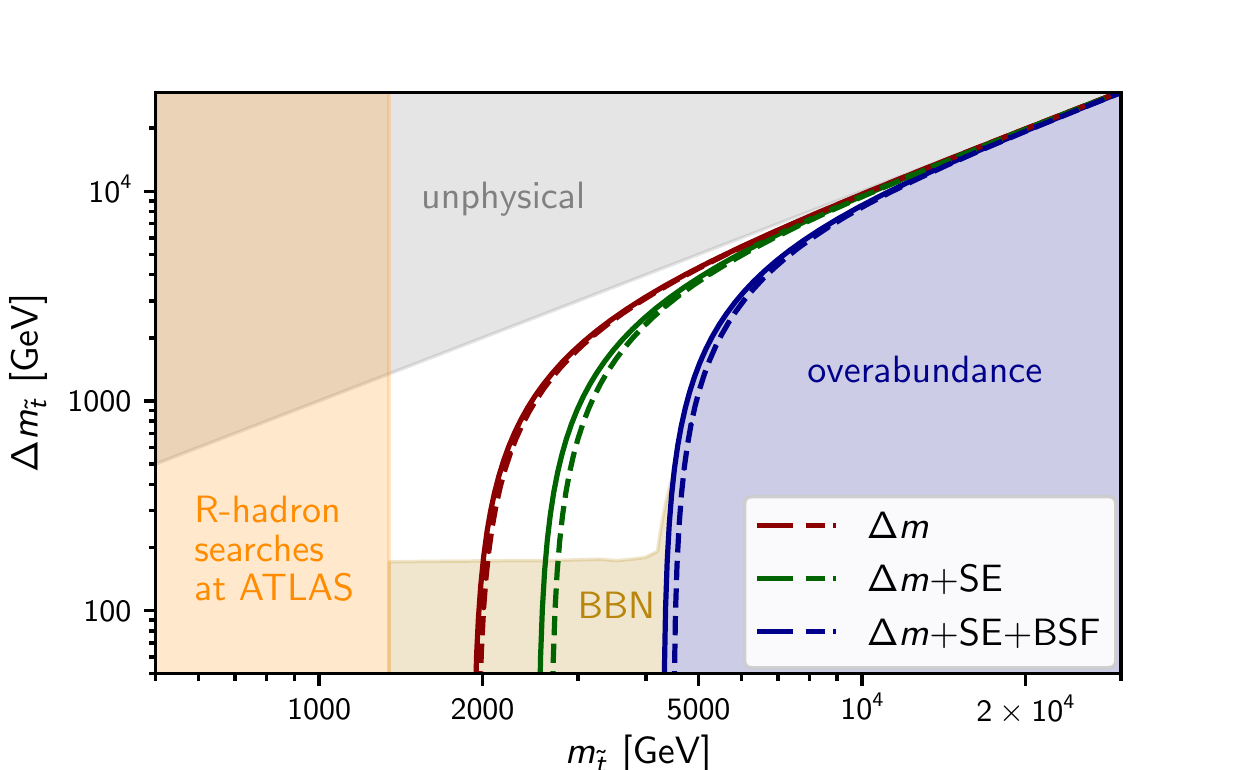} \label{subfig:parameterspaceQCD}}\hspace{0.2cm}
    \subfigure[Viable parameter space for a lepto-philic mediator model assuming $\lH=0$ (solid lines) and $\lH=0.3$ (dashed lines).]{\includegraphics[width=0.75\textwidth]{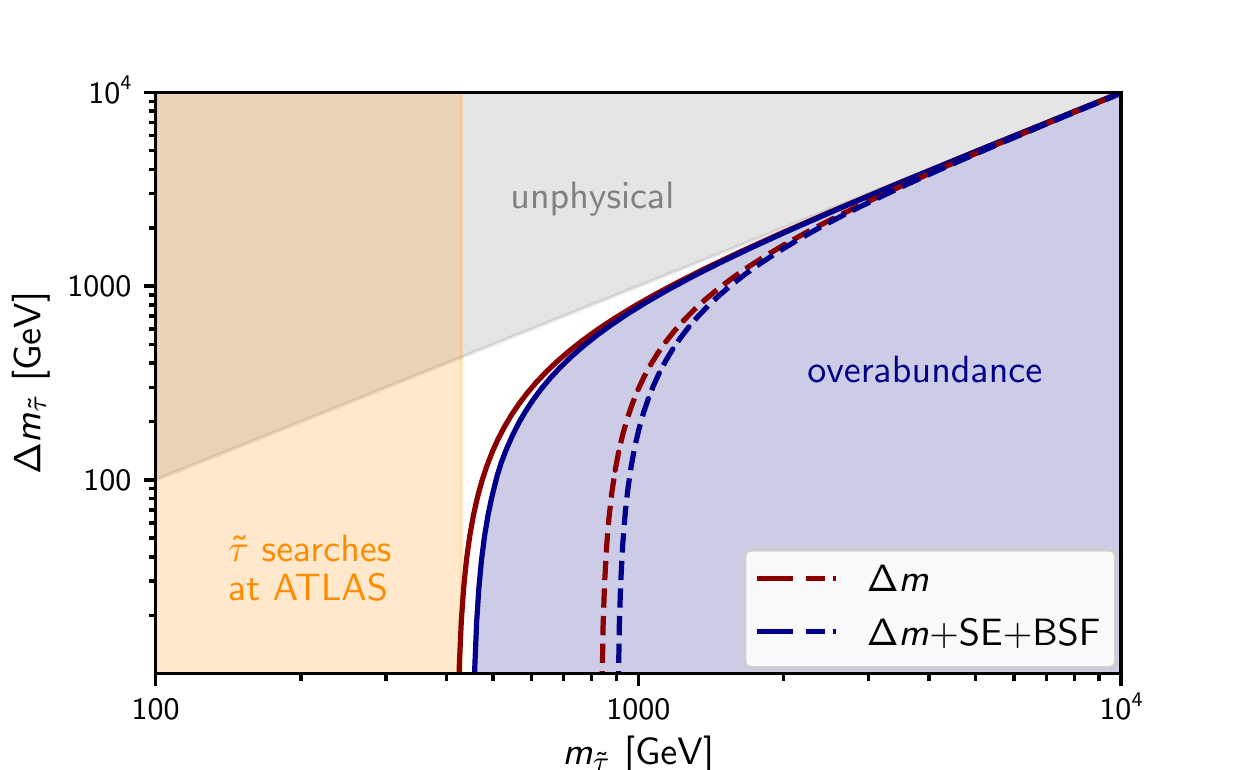} \label{subfig:parameterspaceQED}}
    \caption{The mass difference $\Delta m$ is plotted against the mediator mass marking the superWIMP regime. For the calculation of $\mchi$ uncorrected (red), SE corrected (green) and SE+BSF corrected (blue) yields of the mediator after freeze-out have been used. Shaded are the regions of overabundance for the SE+BSF corrected case (blue) as well as unphysical regions (gray) and regions that have been already excluded by ATLAS searches (orange) or cosmological constraints (yellow). The accessible parameter space considering non-perturbative corrections in the model is left blank, whilst $\lambda_\chi$ is fixed via the total DM abundance $\Omega_{\text{\tiny DM}}$.}
    \label{fig:parameterspace}
\end{figure}

\begin{figure*}
    \subfigure[$\lchi$ (left) as well as $p$ (right) heatmap for the parameter space of a top-philic mediator.]{\includegraphics[width=0.5\textwidth]{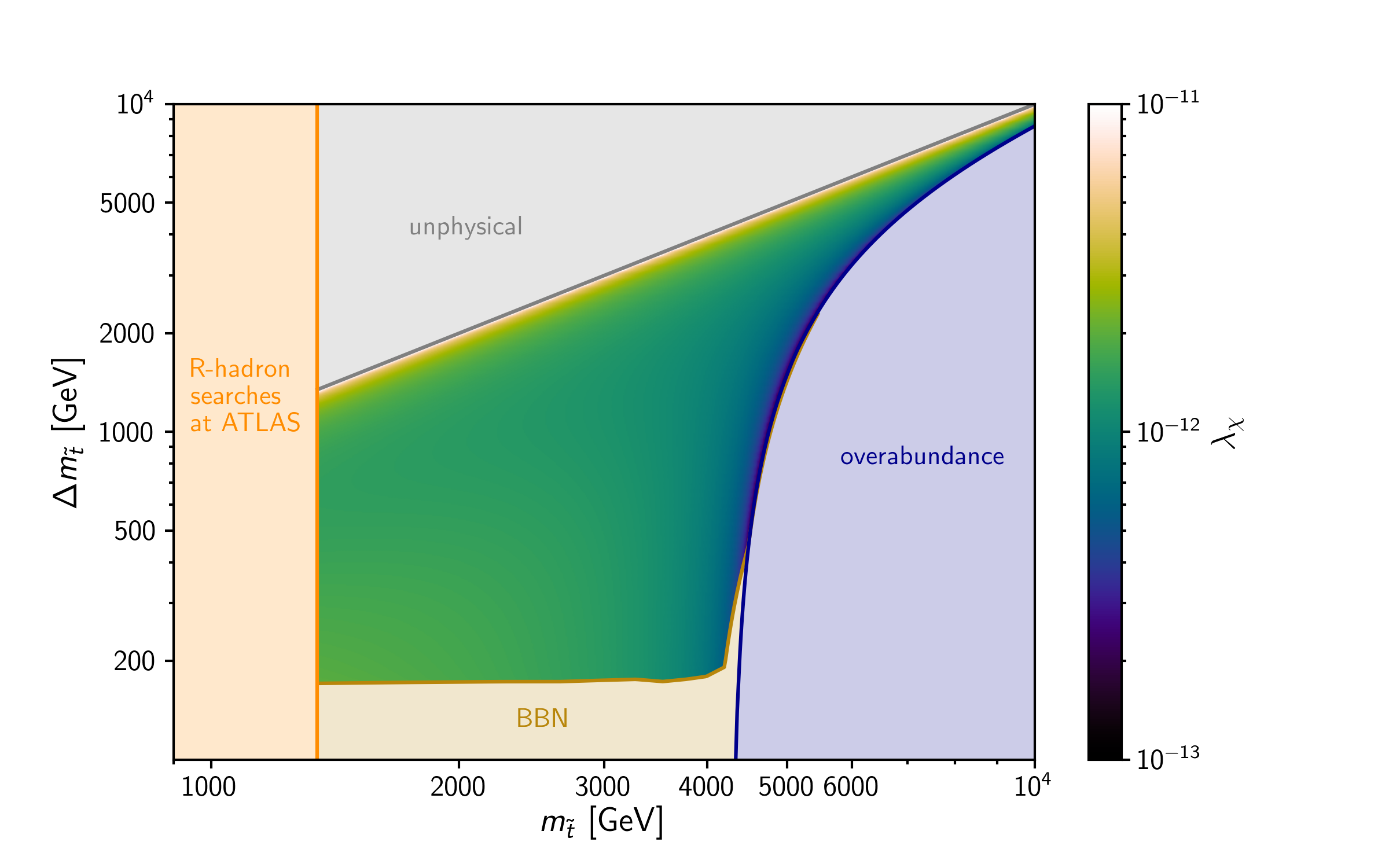} \includegraphics[width=0.5\textwidth]{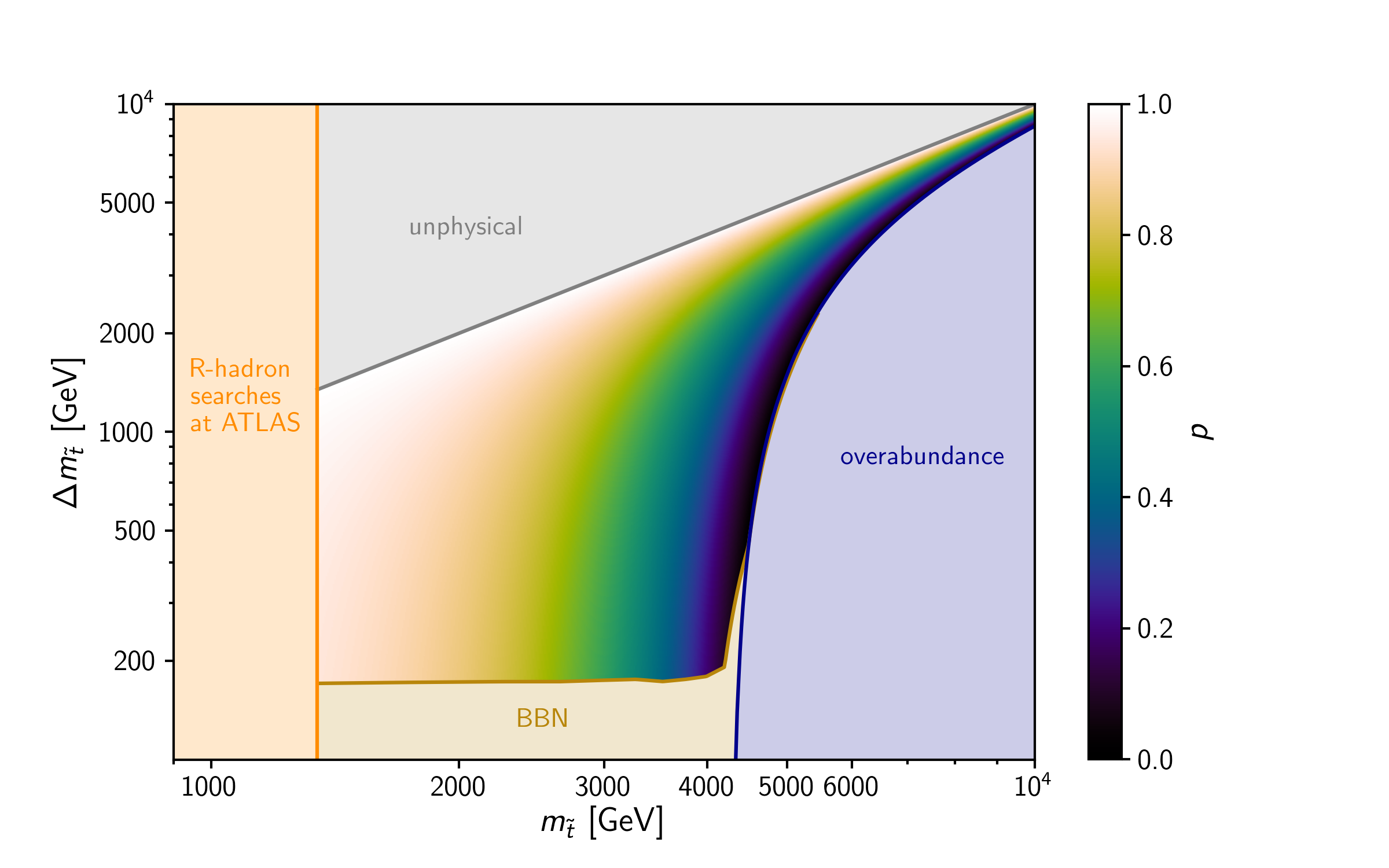} \label{subfig:heatplotQCD}}
    \subfigure[$\lchi$ (left) as well as $p$ (right) heatmap for the parameter space of a lepto-philic mediator.]{\includegraphics[width=0.5\textwidth]{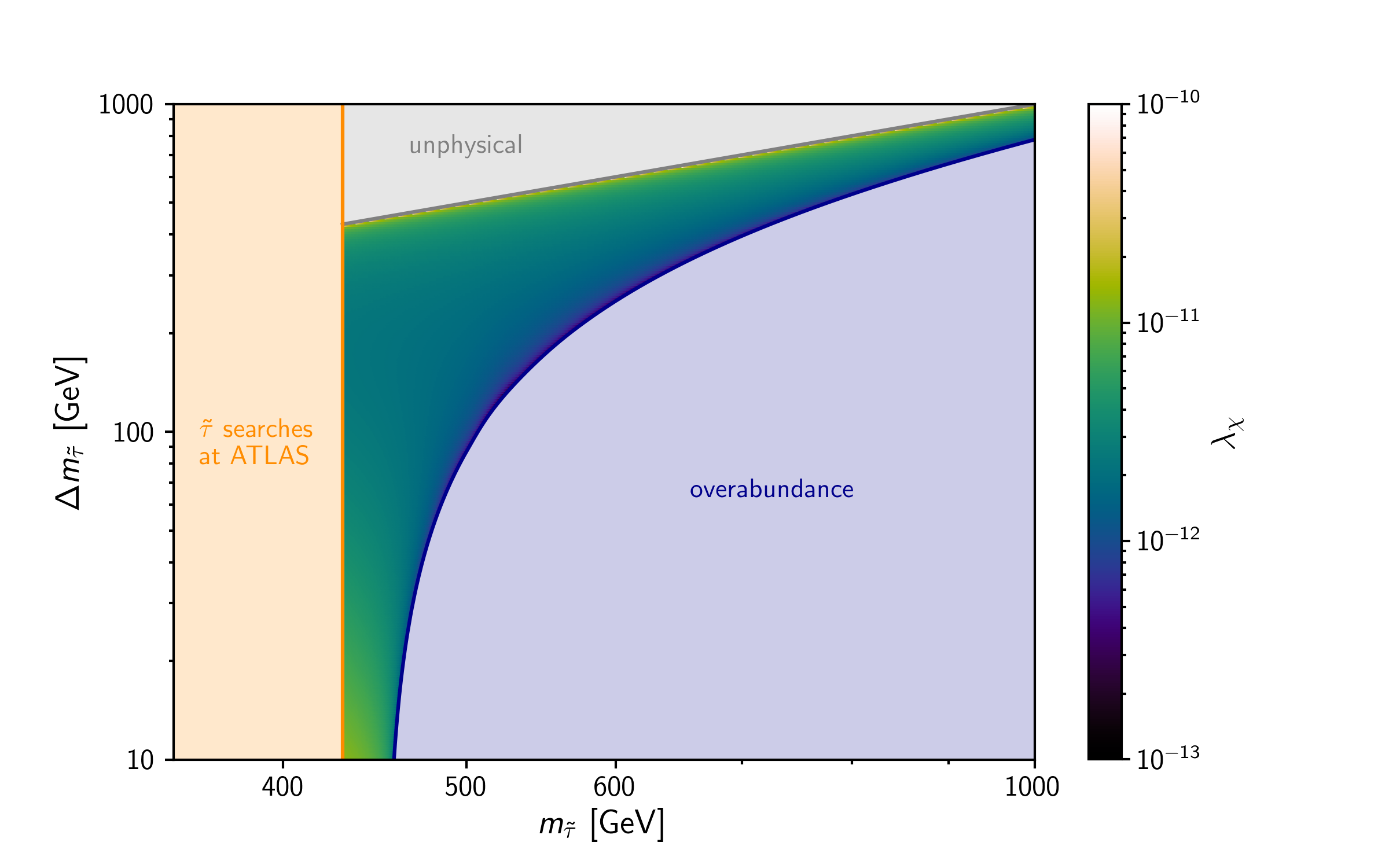} \includegraphics[width=0.5\textwidth]{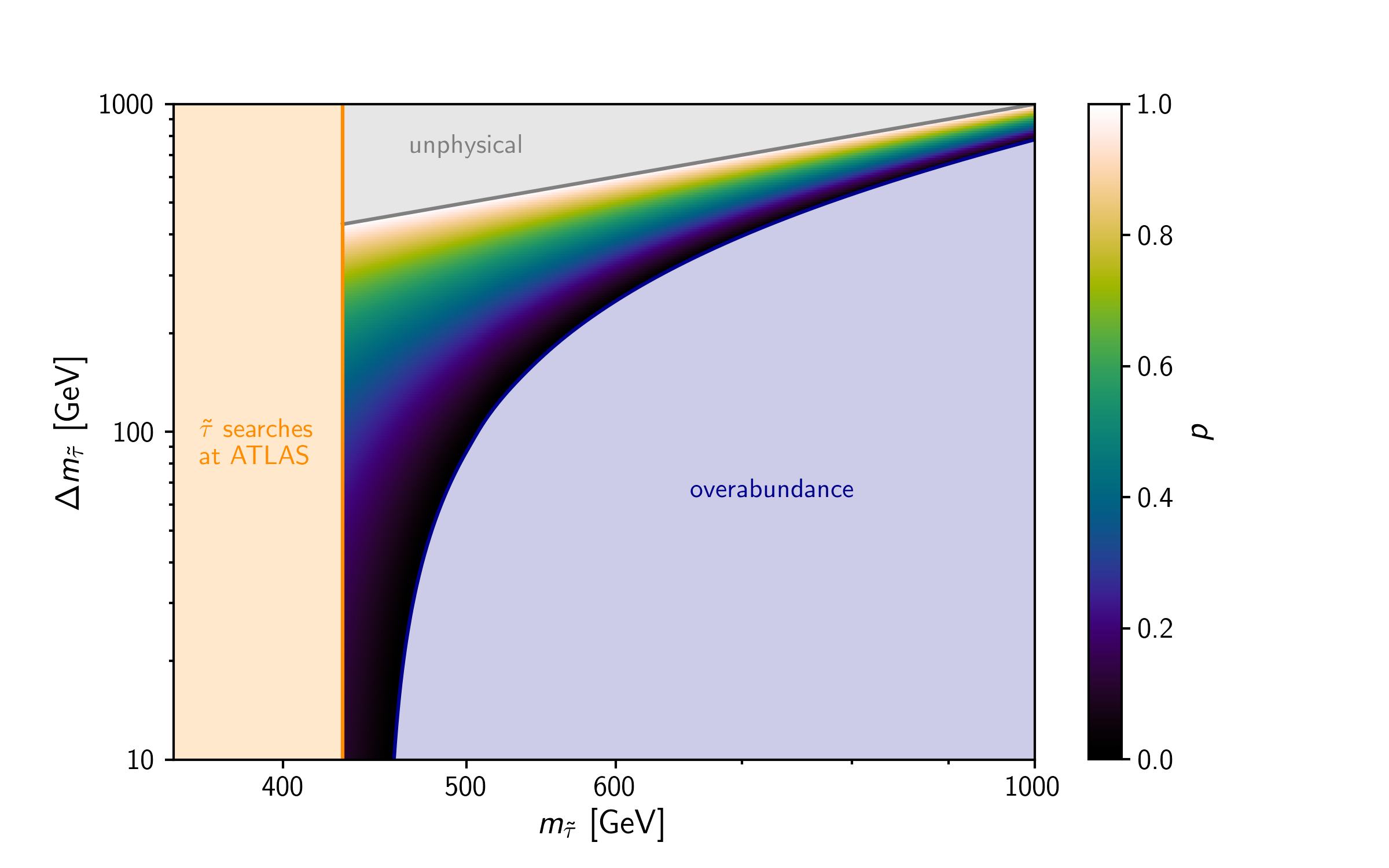} \label{subfig:heatplotQED}}
    \caption{The Yukawa coupling strength $\lchi$ (left) calculated at each point with Eq.~\ref{eq:lambdachi} as well as the percentage of the freeze-in contribution $p=1-\Omega_{\text{sW}}/\Omega_{\text{\tiny DM}}$ (right) in the accessible regime of the parameter space are displayed considering SE and BSF effects for $\lambda_H=0$.}
    \label{fig:heatplots}
\end{figure*}

We illustrate our result in Fig.~\ref{fig:parameterspace} and specify our predictions on $\lambda_\chi$ and the FI fraction $p$ in Fig.~\ref{fig:heatplots}. Since we are chiefly interested in the impact of LHC searches we show the parameter space spanned by the mediator mass and the mass difference $\deltamttilde=\mttilde-\mchi$. The parameter space that allows for successful non-thermal production is bounded by different constraints. 
From above, we are limited by the trivial condition that $\deltamttilde \leq \mttilde$. In principle, there is a small excluded region immediately next to this line which comes from the Lyman-$\alpha$ bound that requires $m_\chi\gtrsim \mathcal{O}(10 \mbox{keV})$ for freeze-in dominated production. However, this is so narrow that it is not visible in this plot. In addition, the estimates in the previous section indicated that there could be a warm dark matter mass bound for a large superWIMP component in this region. 
As already mentioned in the previous section, in the region of parameter space where DM is dominantly superWIMP produced is subject a to lower bound on thee coupling $\lambda_\chi\gtrsim \SI{1.5e-14}{}$ for mediator masses of $\mttilde\sim\SI{50}{\TeV}$ and $\lambda_\chi\gtrsim \SI{e-15}{}$ for $\mttilde\sim\SI{5}{\TeV}$. As can be seen in Fig.~\ref{fig:heatplots} this is not very constraining and an essentially pure superWIMP scenario is allowed for the masses considered here. 
From the left, LHC searches for heavy stable particles are cutting off the allowed parameter space. Since the decay length of the mediator is sufficient to ensure that the bulk of the produced particles does not decay inside the detector, this bound is independent of the DM mass and shows up as a straight line at the lowest $\mttilde$ ($\mtautilde$) which is allowed by observations. 
On the other side, there is a limit from cosmological DM production. At large mediator masses, the contribution from the superWIMP mechanism to the DM abundance exceeds the observed value unless the DM particle is light (we refer to this as an overabundance constraint). 
This pushes for higher values of  $\deltamttilde$. To the left of this limit, the superWIMP mechanism underproduces DM such that freeze-in can make up for the rest, which is emphasized by both plots on the right of Fig.~\ref{fig:heatplots}.  
As can be seen, non-perturbative corrections have a sizable effect on this limit. For the case of a color-charged mediator as shown in Fig.\ref{subfig:parameterspaceQCD}, the allowed $\mttilde$  from the perturbative results at low $\deltamttilde$ of about $\SI{2}{\TeV}$ increases to more than $\SI{4}{\TeV}$ if both bound state formation and the Sommerfeld effect are taken into account. Neglecting bound states leads to an intermediate result that is closer to the perturbative expectation. The impact of $\lH$ is not so pronounced here but still allows a shift of the border of the allowed parameter space by up to $200$ GeV. 

The situation is quite different for the lepto-philic mediator. As expected, the overall smaller cross sections in this scenario push the regime in which DM is overproduced by the superWIMP contribution alone to lower mediator masses while the relative importance of non-perturbative effects and $\lH$ is reversed. Bound state formation has only a small impact here and the results that just include the Sommerfeld effect cannot be distinguished from the full results on the scale of Fig.~\ref{subfig:parameterspaceQED}. It is interesting to note that the current ALTAS limits are already very close to the cosmological upper limit for small  $\deltamtautilde=\mtautilde-\mchi$. However, the Higgs coupling has a much more pronounced impact here and can significantly extend the parameter space for a successful freeze-in in this scenario. 

Finally, BBN leads to a limit that is almost independent of $\mttilde$ and excludes $\deltamttilde \lesssim m_t$. Around this value, the two-body decay of the mediator becomes kinematically inaccessible which leads to a drastic increase of the lifetime for a constant coupling. This is partially compensated for by the fact that also freeze-in from decays becomes inefficient which leads to the relic density constraint favoring larger $\lambda_\chi$ (this effect is also visible in the left plots of Fig.~\ref{fig:heatplots}). However, due to the freeze-in contribution from scattering, which is not affected strongly by the two-body decay threshold, this growth is not strong enough to compensate for the suppression from the three-body decay. Very close to the overproduction bound the BBN limits also get strengthened and deviate from the top mass. Here, most of DM is produced by the superWIMP mechanism and only a smallish fraction is due to freeze-in. Therefore, smaller values for $\lambda_\chi$ are preferred in this region which boosts the lifetime of $\tilde{t}$ and makes the BBN bound stronger. In principle, a similar bound exists for $\tautilde$. However, here the kinematic threshold is only reached at $\deltamtautilde= m_\tau$ and is thus not visible in Fig.~\ref{subfig:parameterspaceQED}.

From the left plot in Fig.~\ref{subfig:heatplotQCD} (Fig.~\ref{subfig:heatplotQED}) where we consider a top-philic (lepto-philic) mediator, we observe that the predictions for $\lambda_\chi$ are relatively constant and of order $ 10^{-12}$ in the regime where both freeze-in and the superWIMP mechanism are sizable. This indicates that the freeze-in and superWIMP regime are well separated in $\lambda_\chi$, such that either of them can be neglected if $\lambda_\chi\ll 10^{-12}$ or $\lambda_\chi\gg 10^{-12}$, respectively. If the mass splitting is very large compared to the mediator mass (which is the case for points on the upper edge of the unconstrained parameter space) freeze-in dominates as can be seen on the right side of Fig.~\ref{subfig:heatplotQCD} (Fig.~\ref{subfig:heatplotQED}). This is especially important, because in this regime the decay temperature of the mediator rises, such that our approach in the superWIMP mechanism to separate the mediator freeze-out from its late decay might no longer be valid. However, since the regime is freeze-in dominated this potential issue has no effect on the DM abundance.

All considered, we find that a perturbative computation significantly underestimates the extend of the cosmologically allowed parameter space for a color-charged mediator. Given that the expected limits on $\mttilde$ will only increase to $\approx \SI{1.6}{\TeV}$ at the HL-LHC makes it hard to test this mechanism for DM production without a collider that operates at a significantly higher center of mass energy than the LHC. In the case of a lepto-philic mediator, the non-perturbative corrections are less important but we find that a computation that assumes $\lH=0$ is not sufficient to establish the border of the non-thermal DM parameter space. In a minimal scenario with a negligible Higgs coupling, the expected improvement of the LHC bound is starting to push the DM towards lighter masses, however, if larger values of $\lH$ are allowed this conclusion can be evaded. Testing this possibility at colliders thus remains challenging.


\section{\label{sec:conclusions}Conclusions}

Non-thermal DM is an attractive possibility which predicts experimental signatures that are radically different compared with DM candidates produced by thermal freeze-out. Studies of other states in the dark sector are potentially very promising in this scenario since the condition for non-thermalization of DM points towards very weak interactions that can easily render these particles long-lived. Particularly interesting in this context are models in which the dark sector partners of the DM possess SM quantum numbers since these allow for a copious production of the new physics at colliders. 

However, it is important to note that the same gauge interactions that are attractive from the point of view of LHC phenomenology also imply large non-perturbative corrections for DM production in the early Universe. Therefore, a realistic assessment of the experimental capabilities calls for a thorough study of the cosmologically preferred parameter space that takes these effects into account. 

We performed a state-of-the-art analysis of non-thermal DM production in a class of minimal simplified models with a fermionic singlet DM candidate and a (color-)charged mediator where we included both, the freeze-in and the superWIMP mechanism and considered corrections from the Sommerfeld effect as well as from bound state formation.

After comparing our results with a perturbative calculation, we observe that for a color-charged mediator neglecting the non-perturbative effects overpredicts the superWIMP contribution to the DM abundance by an order of magnitude. It is clear that such a large correction has to be taken into account and we find that the parameter space that allows for non-thermal DM production is significantly larger than previously anticipated. In the case of a lepto-philic mediator, the non-perturbative effects are much more modest and only amount to an $\mathcal{O}(20\%)$ correction. Note, however, that this case is very sensitive to the coupling between the mediator and the Higgs. Considering a modest value of $\lambda_H=0.3$ instead of the minimal choice of $\lambda_H=0$ leads to a change in the superWIMP contribution to $\Omega_{\mbox{\tiny DM}}$ of up to one order of magnitude. Consequently, a sizable extension of the parameter space that allows for non-thermal production is also possible here albeit through a different effect.

Combining bounds from LHC searches, cosmological constraints from BBN, and the predictions from DM production in the early Universe, we have analyzed both models. While for comparably small mediator masses of a few $\SI{}{\TeV}$ also a small mass gap between DM and the mediator is possible, for larger masses it is inevitable to have a large mass separation to avoid an overabundance of DM in the universe. Non-perturbative effects have been shown to broaden this parameter space significantly. 
Overall, our results show that collider tests of these models are more challenging than expected. Adding more luminosity at the HL-LHC will extend the reach of the experimental search. However, a collider with a larger center of mass energy would be very beneficial for testing these models of DM.


\section*{Note added}
\noindent During the completion of this manuscript, the preprint \cite{Decant:2021mhj} appeared on the arXiv which also considers the impact of bound state formation on the superWIMP mechanism in a model with a top-philic mediator. Their study focuses on astrophysical limits on non-thermal DM and is complementary to our analysis.


\section*{Acknowledgements}
\noindent We acknowledge the support of the Research Training Group RTG2044 funded by the German Research Foundation (DFG).


\appendix
\appendixpage
\renewcommand{\theequation}{\thesection.\arabic{equation}}


\section{\label{app:freezein}Freeze-in yield for a color-charged mediator}
For freeze-in, the Boltzmann equation for the yield can be solved by direct integration. The two contributions to Eq.~\ref{eq:freezein} describe the $1\to2$ particle decay $\ttilde\to t_R+\chi$ as well as the $2\to2$ processes which are relevant in the considered regime, namely $\ttilde+\bar{t}_R\to g+\chi$, $\ttilde+g\to t_R+\chi$ and $t_R+g\to \ttilde+\chi$. The two contributions are given by
\begin{align}
    \label{eq:Yfi1to2}
	Y^{\text{FI}}_{\chi,1\to 2}(x)=&\xi_1^{\text{FI}}\Gamma_{\ttilde}\int_0^x\dd{x'}\frac{\gstar^{1/2}(x')}{\heff(x')}Y^{eq}_{\ttilde}(x') x'\frac{\kone{x'}}{\ktwo{x'}}\qquad\\
	Y^{\text{FI}}_{\chi,2\to 2}(x)=&\xi_2^{\text{FI}}\int_0^x \dd{x'}\gstar^{1/2}(x')\frac{Y^{eq}_a(x')Y^{eq}_b(x')}{x'^2}\expval{\sigma \vrel}\label{eq:Yfi2to2}
\end{align}
 with prefactors $\xi_1^{\text{FI}}=\sqrt{45/(4\pi^3)}M_{\text{Pl}}/\mttilde^2$, $\xi_2^{\text{FI}}=\sqrt{\pi/45}M_{\text{Pl}}\mttilde$ and $a,b$ denoting the incoming particles in the $2\to 2$ processes. The $1\to 2$ decay rate $\Gamma_{\ttilde\to t_R\chi}$ reads
\begin{eqnarray}
    \Gamma_{\ttilde\to t_R\chi}&=& 
    \lchi^2\frac{\sqrt{\lambda(m_{\tilde{t}}^2,m_t^2,m_\chi^2)} (\mttilde^2-\mchi^2-m_t^2)}{16\pi\mttilde^3}\label{eq:decayrate}\quad
\end{eqnarray}
where $\lambda(x,y,z)$ is the standard K\"all$\acute{\mbox{e}}$n function. The equilibrium yield $Y^{eq}_{i}(x)=n^{eq}_i(x)/s(x)$ for massive and massless particles following the Boltzmann statistics is given by $Y^{eq}_{i}(x)=45g_i\eta(m_i,x)/ (2\pi^4\mttilde^3\heff(x))$ with 
\begin{equation}
    \eta(m_i,x)=
\begin{cases}
   \frac{1}{2}\mttilde m_i^2x^2\ktwo{\frac{m_i}{\mttilde}x}\quad&\text{for}\quad m_i\neq 0\\
   \mttilde^3\quad&\text{for}\quad m_i=0
\end{cases}
\end{equation}
where $g_i$ are the internal degrees of freedom of particle species $i$ and $\heff$ is the effective number of entropy degrees of freedom. The parameter $\gstar$ is defined as \cite{Gondolo1991}
\begin{equation}
    \gstar^{1/2}=\frac{\heff}{\geff^{1/2}}\left(1+\frac{1}{3}\frac{T}{\heff}\dv{\heff}{T}\right)\,,
\end{equation}
and depends on the effective number of energy and entropy degrees of freedom $\geff$ and $\heff$. Temperature dependent results for them can be found in e.g. \cite{Husdal2016}. For generic $2\to2$ processes, the thermally averaged cross section is given by \cite{Edsjo:1997bg}
\begin{eqnarray}
	\expval{\sigma \vrel}&=&\frac{\mttilde x^5}{32\eta(m_a,x) \eta(m_b,x)}\int_{s_{\text{min}}}^\infty\dd{s}\frac{\lambda(s,m_a^2,m_b^2)}{\sqrt{s}}\nonumber\\
	&&\kone{\sqrt{s}\frac{x}{\mttilde}}\sigma_{ab\to cd}(s)
\end{eqnarray}
where $s=(p_a+p_b)^2$ and  $\sigma_{ab\to cd}(s)$ denote the usual center of mass energy and the cross section. If the temperature dependence of $\geff$ and $\heff$ is neglected, \ref{eq:Yfi1to2} can be integrated analytically. We always take the full temperature dependence into account and integrate both Eq.~\ref{eq:Yfi1to2} and Eq.~\ref{eq:Yfi2to2} numerically. 

The integration over the Mandelstam variable $s$ in the $\ttilde + g \to t_R + \chi$ channel requires some care since it is IR divergent at tree level. Including loop diagrams to cancel the divergence goes beyond the scope of this work and we follow \cite{Garny2018} by setting a minimal bound of $\sqrt{s_{\text{min}}}=(1+\epsilon) \mttilde$ with $\epsilon=0.1.$ instead.


\section{\label{app:bsfmassbound}Potential strengths and bound state formation mass bound}
\begin{table}
    \centering
    \caption{Attractive potentials and fine structure constants for color-charged $\ttilde-\ttilde^*$ mediator interactions ($Q_{\text{em}}=2/3$). For lepto-philic $\tautilde-\tautilde^*$ mediator interactions, the results for $\gamma$, $Z$ and $H$ exchange remain the same with $Q_{\text{em}}=-1$.}
    \label{tab:potentials}
    \begin{tabular}{ccc}
        \toprule
		gauge boson & $V(r)$ &$\alpha$ \\
		\midrule
		Gluon & $V_g(r)=-\frac{\alpha_{g,[\mathbf{1}]}}{r}$ & $\alpha_{g,[\mathbf{1}]}=\frac{4}{3}\alpha_s$\\[0.5cm]
		Photon & $V_\gamma(r)=-\frac{\alpha_\gamma}{r}$ & $\alpha_\gamma=Q_{\text{em}}^2\alphaem$\\[0.5cm]
		Z-boson & $V_Z(r)=-\frac{\alpha_Z}{r} e^{-m_Zr}$ & $\alpha_Z=Q_{\text{em}}^2\tan^2\theta_W\alphaem$\\[0.5cm]
		Higgs & $V_H(r)=-\frac{\alpha_H}{r} e^{-m_Hr}$& $\alpha_H =\frac{\lH^2v^2}{16\pi \mttilde^2}$\\
		\bottomrule
    \end{tabular}
\end{table}
In order to determine the importance of the non-perturbative corrections coming from different interactions, we have summarized the quantum mechanical potentials as well as their interaction strengths in Table \ref{tab:potentials}.

Bound states can only form in sufficiently long-ranged potentials. In a hot plasma with temperatures above the confinement scale, the gluonic interactions lead to a Coulomb potential. These are always long-ranged and can support bound states. For color-charged mediators, the gluon potential dominates and we can safely neglect contributions from all other bound states\footnote{The Higgs potential of our $SU(2)_L$ singlet mediator cannot compete with the gluon potential in the allowed mass range even if values up to the perturbative limit $\lH\leq \sqrt{4\pi}$ are allowed. This can be circumvented in theories with additional sources of electroweak symmetry breaking that allow for a three-point interaction between the color-charged mediator and the Higgs as considered in \cite{Harz:2017dlj,Harz:2019rro}.}. In the case of a lepto-philic mediator, also contributions from $Z$ and $H$ boson exchange could play a role in bound state formation. These are described by Yukawa potentials, for which it is less obvious if they are sufficiently long-ranged in the energy regime considered. This can be answered by demanding that the screening length $D=1/m_X$ has to be above the critical value that marks the point of zero binding energy. Numerical studies showed $ D_0 \geq 0.8399 \, a_0$ is required for a $n=1$ bound state to exist where $a_0=2/(\alpha\mttilde)$ is the Bohr radius \cite{Rogers1970}. Thus, we obtain the limit on the model parameters that support a pure Yukawa bound state
\begin{equation}
	\mttilde\geq\frac{1.68 m_X}{\alpha}.
\end{equation}
For the Z exchange $\alpha_Z=0.00224$ (for $Q_{\text{em}}=-1$) we get $\mttilde\gtrsim \SI{68}{\TeV}$ which is well outside of our mass region of interest. In the case of Higgs exchange, $\alpha_H$ depends on the mediator mass and, thus, even for sizable $\lH\sim\order{1}$, the strength of the potential is too weak to allow for a bound state in the regime where $m_H \leq \mttilde$. Similar reasoning holds for the lepto-philic mediators. Therefore, the Coulomb potentials of the gluon and photon, respectively, will play the lead role in the bound state. Note, however, that this does not imply that the $H$ and $Z$ contributions will not affect the Sommerfeld factors and we take them into account there.


\section{\label{app:QEDcase} The simplified model with a lepto-philic mediator}
The DS and interaction Lagrangian for a colorless mediator $\tautilde\subset(\mathbf{1},\mathbf{1},-2)$ (still considering the same $\chi$) reads
\begin{eqnarray}
	\mathcal{L}_{\text{DS}}&=& i\bar{\chi}\gamma^\mu\partial_\mu\chi-\frac{1}{2}\mchi^2\bar{\chi}\chi-\mtautilde^2\tautilde^*\tautilde\\
	\mathcal{L}_{\text{int}}&=&\abs{D_\mu \tautilde}^2+\lH \tautilde\,\tautilde^*\abs{\Phi}^2+\lchi \overline{\tau}_R\tautilde\chi+h.c.
\end{eqnarray}
where we couple to a right-handed $\tau_R$. 

The freeze-in yield $Y^{\text{FI}}_\chi(x)\approx2Y^{\text{FI}}_{\chi,\tautilde\to \tau_R\chi}(x)$ in this case, since the $\alphaem^2$ dependence of the scattering processes makes them inefficient such that they do not contribute significantly for $\mttilde -m_\chi \gtrsim m_\tau$. In the relevant expressions we just need to adjust the internal degrees of freedom $g_{\ttilde}\to g_{\tautilde}=1$ and the particle masses. For the DM production process through the superWIMP mechanism, Eq.~\ref{eq:BSFBoltzmannAll} remains qualitatively the same. However, the cross sections and rates change. Since we can neglect the $H$ and $Z$ contributions for BSF processes (see Appx.~\ref{app:bsfmassbound}), the modifications to Eq.~\ref{eq:avgBSFcrossection} are rather simple. One replaces all QCD couplings $\{\alphasBs,\alphagBs,\alphasBSFs\}\to\alpha_\gamma\equiv Q_{\text{em}}^2\alphaem(m_Z)$ (also in Eq.~\ref{eq:YBeq}) and removes the pre-factors arising from an $SU(3)_c$ charge. The BSF cross section for $\tautilde + \tautilde^* \to \mathcal{B}(\tautilde,\tautilde^*) +\gamma$ then reads
\begin{equation}
    \sigma_{\text{BSF}}\vrel\approx\frac{2^9}{3}\frac{\pi\alpha_\gamma^2}{\mtautilde^2}S_0(\zeta_\gamma)\frac{\zeta_\gamma^4e^{-4\zeta_\gamma\arccot(\zeta_\gamma)}}{(1+\zeta_\gamma^2)^2}
\end{equation}
with $\zeta_\gamma=\alpha_\gamma/\vrel$.

For the s-wave annihilation cross section of the $\tautilde + \tautilde^*$ state  we consider $\gamma+\gamma$, $\gamma+Z$, $Z+Z$, $H+H$, $W^++W^-$ and $t+\bar{t}$ (for $\lH=0$ only the first three) as the most prominent final states. In order to calculate $\expval{S_{\text{ann},0}}$ we need to implement a numerical solution due to the non-negligible contributions from  $Z$ and $H$ boson exchange to the Sommerfeld factor. Following the approach of \cite{Iengo2009} we solve the differential equation 
\begin{equation}
	\varphi''(y)+\left(1+\frac{2}{y}\left[\zeta_\gamma+\zeta_Ze^{-b_Zy}+\zeta_He^{-b_Hy}\right]\right)\varphi(y)=0
\end{equation}
which can be derived from Eq.~\ref{eq:SEscattering} by substitution $\phik^{0}(\vec{r})=\varphi(y)/(Cy)$, $y=kr$, $\zeta_i=\alpha_i/\vrel$ with couplings given in Table \ref{tab:potentials} and $b_i=2m_i/(\mtautilde\vrel)$. We choose $\lim_{y\to 0}\varphi(y)=y$ and $\lim_{y\to 0}\varphi'(y)=1-(\zeta_\gamma+\zeta_Z+\zeta_H)y$ as initial conditions such that we can extract $C^2=\lim_{y\to\infty}[\varphi^2(y)+\varphi^2(y-\pi/2)]$ from asymptotic behaviour of $\lim_{y\to\infty}\varphi(y)=C\sin(y)$. This gives us $\expval{S_{\text{ann},0}}$ from Eq.~\ref{eq:Sann0Ri} considering  $S_{\text{ann},0}=\abs{\phik^{0}(0)}^2=1/C^2$.
Finally, the lowest order decay rate for a lepto-philic mediator bound state is given by 
\begin{equation}
    \Gamma_{\text{dec}}\approx \frac{g_{\tautilde}^2}{g_B}\frac{\mtautilde^3\alpha_\gamma^3}{8\pi}\sum_i \sigma^i_0.
\end{equation}



\bibliographystyle{hieeetr}
\bibliography{references.bib}

\end{document}